\documentclass[letterpaper, 10 pt, conference]{ieeeconf}  

\IEEEoverridecommandlockouts                              
\usepackage{graphicx}
\usepackage{morefloats}
\usepackage{subfig}
\usepackage{upgreek}
\usepackage{amssymb}
\usepackage{amsmath}

\title{\LARGE \bf
EigenRank by Committee: A Data Subset Selection and Failure Prediction paradigm for Robust Deep Learning based Medical Image Segmentation
}

\author{Bilwaj Gaonkar\textsuperscript{a}, Joel Beckett \textsuperscript{a}, Mark Attiah\textsuperscript{a}, Christine Ahn\textsuperscript{a}, Matthew Edwards \textsuperscript{a},  Bayard Wilson\textsuperscript{a}, \\
Azim Laiwalla\textsuperscript{a} , Banafsheh Salehi\textsuperscript{b}, 
Bryan Yoo\textsuperscript{b},  Alex Bui\textsuperscript{b}, Luke Macyszyn \textsuperscript{a}\\ \\
\textsuperscript{a} \textit{Department of Neurosurgery, UCLA}\\
\textsuperscript{b} \textit{Department of Radiological Sciences, UCLA}\\
}

\begin{document}

\maketitle
\thispagestyle{empty}
\pagestyle{empty}

\begin{abstract}
Translation of fully automated deep learning based medical image segmentation technologies to clinical workflows face two main algorithmic challenges. The first, is the collection and archival of large quantities of manually annotated ground truth data for both training and validation. The second is the relative inability of the majority of deep learning based segmentation techniques to alert physicians to a likely segmentation failure. Here we propose a novel algorithm, named `Eigenrank' which addresses  both of these challenges. Eigenrank can select for manual labeling, a subset of medical images from a large database,  such that a U-Net trained on this subset is superior to one trained on a randomly selected subset of the same size. Eigenrank can also be used to pick out,  cases in a large database, where deep learning segmentation will fail. We present our algorithm, followed by results and a discussion of how Eigenrank exploits the Von Neumann information to perform both data subset selection and failure prediction for medical image segmentation using deep learning.
\end{abstract}

\section{INTRODUCTION}
\subsection{Significance}
Deep learning methods have become the mainstay of fully automatic medical image segmentation. These methods play a key role in the  development of quantitative imaging biomarkers for a number of pathologies. However, training and deploying deep learning segmentation in practice is beset by a number of challenges. Two significant but related challenges are:
\begin{itemize}
\item Data subset selection (DSS) -  the development of robust segmentation tools by using human annotation efforts in the most efficient possible manner
\item Failure Prediction (FP)  - the ability to predict on which cases a deep learning based segmentation model will fail.
\end{itemize}
Both problems are  significant in medical image segmentation, more than natural image segmentation. This is because, availability of expert annotated data for training medical image segmentation models is severely constrained. These models need to be robust to natural and pathologic variation, despite training on datasets much smaller than those regularly used in natural image segmentation competitions. While, standard  machine learning philosophy, improves robustness by training on increasingly larger sets of training data, the cost of annotation is much higher in medical imaging. Given the limited availability of physician effort, it is important that manual annotation efforts be utilized in the most efficient manner, when creating a new training set aimed at segmenting a specific anatomical region. Thus,  we must be able to optimally choose a training subset of images for manual annotation from within the vast store of imaging data available in a standard hospital PACS. Moreover, this subset must be selected, without the availability of any manual segmentations on the PACS. This is the data subset selection (DSS) problem of medical image segmentation, which we address in this work.

Failure prediction is a problem that emerges when one attempts to incorporate automatic medical image segmentation algorithms into clinical workflows. An algorithmic framework is not expected to be perfect. However, an algorithm which is imperfect, and can alert the attending physician to its imperfections is far more valuable than an algorithm which fails silently. The majority of existing algorithms for medical image segmentation fail silently. In this work, we have developed a framework which can  \textit{pick out} scans where a deep U-Net algorithm will  fail. We expect that addressing failure prediction will be critical to enable any deployment of deep learning segmentation algorithms for clinical imaging.

\subsection{Related Work}
\subsubsection{Data subset selection and Active Learning}
 Typical DSS aims to choose a training  subset from a large dataset, 
 such that  models trained on the subset incur minimal loss compared to models trained on the complete dataset. \cite{wei2015submodularity,schreiber2019apricot}. Active learning on the other hand involved the ability to interactively query the user during the training process \cite{settles2009active}.  DSS and active learning have been a part of machine learning literature, for more than three decades often under  alternate and related headings\cite{settles2009active,rubens2015active,das2016incorporating, zhou2017brief}. Consequently, there exists substantial literature on data subset selection, active learning,  as well as weakly supervised learning, all of which cannot be reviewed here. However, we note that the majority of standard DSS algorithms are designed to work  with binary classification and focus on preserving classification  accuracy. The closest work to ours in literature comes from pathology \cite{yang2017suggestive,di2019deep} where uncertainty at the voxel level is used to trigger a query to the human expert to segment a patch. This strategy of using voxel level disagreement to drive human annotator attention to specific regions of images has also been used with deep ensembles constructed by bootstrap sample selection \cite{dolz2017deep,deng2018strategy}.  A disciplined framework which defines manual annotation minimization as a linear program is described by Bhalgat \cite{bhalgat2018annotation}.  Authors suggest that mixed supervision where weak annotation using landmarks and bounding boxes is combined with relatively few full annotations could be used to improve segmentation quality.  Authors define an active learning based  semi-automatic segmentation technique using Fisher information to optimize manual segmentation effort to differentiate tissue type in infant brains \cite{sourati2018active}. Our method is similar in that, it is based on a Von Neumann information paradigm, but different in the sense that we operate at a whole scan level. Tangentially related work includes \cite{zhao2018deep,gaonkar2016deep}, where multi-level networks are used with one stage detecting a bounding contour while the second stage segments. While these approaches are neither DSS nor active learning, they do reduce the amount of human effort needed for segmentation. The aforementioned methods have mainly been designed to improve semi-automatic segmentation and improve the throughput of manual segmentation. Hence, the aforementioned literature  aims to alleviate manual work by focus on problematic regions, by active learning  at the pixel/voxel level.  In contrast  we approach subset selection at the subject/patient level rather than a  pixel or a  patch level. Our work defines and measures uncertainty between segmentations produced by multiple models at a subject level. The driving motivation in this work is to  make automated segmentation based biomarkers a part of the radiological workflow, where majority of the work may be done by the automation, while identifying cases which will need human attention in the clinic, and then using such cases to improve the automation itself. A second aspect which is not addressed widely in previous literature is that of `robustness'. If clinical workflow automation is the goal, robustness is as important as accuracy. A method which performs consistently, albeit at a slightly lower accuracy is perhaps more valuable than a method which segments at a high `average' accuracy but is less robust. Our approach selects subsets which lead to the creation of deep learning models which are both more accurate and more robust than random selection. We study subset selection from a robustness point of view as opposed to an accuracy point of view. This is another philosophical difference between current art and the work proposed here.

\subsubsection{Failure Prediction}
Failure is a topic of research which has gained wide-attention in deep learning as well as machine learning. Deep learning systems based on convolutional networks, which can attain human level performance on narrow tasks. Yet,  seem to  fail due to incomprehensible reasons, while maintaining `high confidence' in the accuracy of prediction\cite{nguyen2015deep,goodfellow2014explaining}. While the creation of contrived examples to incite algorithmic failure has gained much attention, very little guidance is available on how one could pick real examples from existing datasets, where a particular algorithm might fail. This is especially true of medical image segmentation, where  precious little has been done to predict failure.  Recent ideas include the use of auxiliary networks to predict quality \cite{devries2018leveraging} to using ensembles by training on different subsets of the same dataset \cite{jungo2019assessing}. These tend to operate under the pretext that image segmentation can be modeled as a pixel level classification or regression operation and bringing to bear standard confidence prediction machinery for failure prediction. But image segmentation is not just pixel classification. In this work we propose an approach, which is based on comparing segmentation generated by an iteratively refined ensemble of segmentation models, where the refinement step is informed by the the degree of disagreement between previous models. The degree of disagreement is quantified on the basis of a Dice score, which is a metric specifically designed to evaluate image segmentation.  The framework presented is extensible to incorporate other `segmentation' specific metrics and addresses the clinically relevant problem of `picking out' scans which might be problematic rather than picking out `pixels' which might be problematic. More importantly, we validate our approach on actual clinical data and demonstrate its effectiveness.

\subsection{Contributions}

The main contribution of this work is to propose a novel  iterative algorithm for data subset selection and failure prediction in the medical image segmentation. Our approach  iteratively selects challenging cases from a large dataset, archives models trained on cases selected in each iteration to generate an ensemble of deep learning models. In the next iteration, challenge cases are selected based on the degree of disagreement between all models. The degree of disagreement is defined by the  maximum eigenvalue of a matrix whose entries are the Dice scores comparing segmentations generated by different models in the ensemble. We discuss how this measure is closely connected to the Von Neumann information metric and validate the proposed algorithm in clinical MRI segmentation tasks related to the spine. In broad strokes, the proposed algorithm can be seen as an extension of the query-by-committee framework \cite{seung1992query}  to medical image segmentation using Von Neumann Information metric.
Using spinal canal and intervertebral disk segmentation on MRI  we validate that our algorithm. Our experiments show that our algorithm:  
 \\
1. Chooses a subset of `challenging' cases for initial training \\
2. Yields trained deep learning models,  more robust and more accurate than models trained using random selection\\
3. Accurately identifies entire scans in the data which are challenging with respect to the defined segmentation task, thus enabling failure prediction \\
Our work presents a fundamentally new way to select training data for creating novel segmentation models using deep learning. It also presents a systematic approach  to identify scans  most likely to require human attention, by preempting algorithmic failure. These are fundamental challenges in medical image segmentation and addressing them makes deep learning based segmentation, both more attractive and more defensible for deployment in clinical workflows.

\begin{figure}
  \centering
  \includegraphics[width=0.3\linewidth]{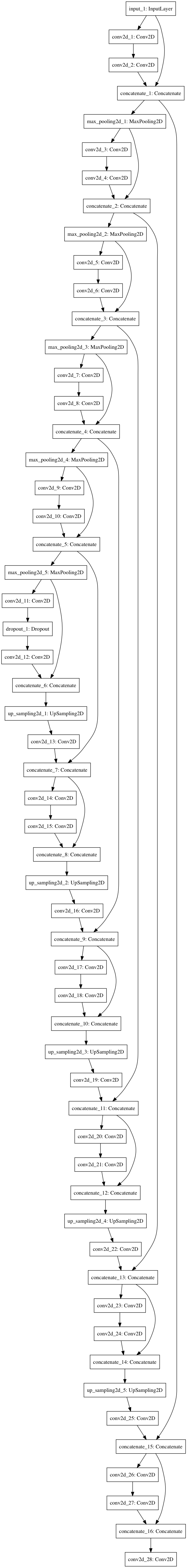}
  \caption{Residual U-Net model used in our experiments}
  \label{fig:model}
\end{figure}


\section{Methods}
The central aim of the investigations presented here is to convince the reader of the value of our novel algorithmic framework for data subset selection and failure prediction in deep learning based medical image segmentation. Normally, large annotated data sets are  thought of as prerequisites for training deep learning methods. \cite{greenspan2016guest}. In this work, we show that data selection using our framework can help create robust and accurate deep learning models with fewer data.  Further, we show that with our algorithmic framework, it is possible to preemptively identify scans, where a deep learning model will fail.

\subsection{Data collection and preprocessing}
 The data used as a part of this work was obtained by  querying the University of California at Los Angeles (UCLA) picture archiving and communication system (PACS) for individuals who had undergone any form of spine imaging using the corresponding CPT (Current Procedural Terminology) codes \cite{Terminology1970} corresponding to lumbar MRI. The search yielded a large number of accession numbers, of which we  selected cases for the purposes of experiments detailed here. This data was obtained under the IRB 16-000196.   Images were downloaded from PACS, anonymized and resampled in the axial or sagittal plane to 256x256-px. Subsequently, each image was converted to the NIFTI \cite{Li:2016kj,Larobina2014} format and linearly histogram matched to a template image using the SimpleITK \cite{lowekamp2013design} package.  Template image intensities were  scaled to lie between a maximum of 1 and a minimum of 0. Linear histogram matching, ensured that the same was true of each image used in this study. 
 
\subsection{Manual Segmentation} 
\begin{figure}
  \centering
  \includegraphics[width=0.95\linewidth]{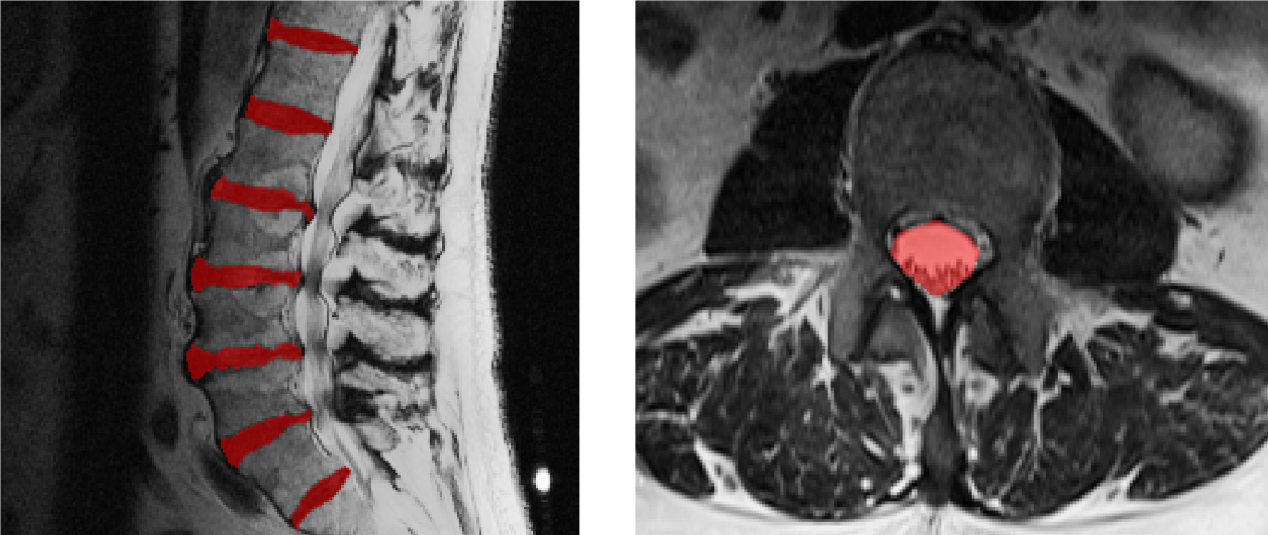}
  \caption{Illustration of intervertebral disk and spinal canal segmentation}
  \label{fig:segexample}
\end{figure}

 Manual segmentation of spinal canals was performed by two medical students using ITK-SNAP \cite{Yushkevich:2006fk} and validated by a attending physician. The manual segmentation data were used as ground truth for all experiments presented here. The tasks we focus on  consists of image segmentation of spinal canals and inter-vertebral disks on MRI data. We have previously published on this task, and enumerated challenges involved in the process. Examples of intervertebral disk segmentations and spinal canal segmentations are shown in figure\ref{fig:segexample}.
 
\subsection{Model Architecture and parameterization}
We use a standard model architecture called the residual U-Net. The U-Net which was first proposed for cellular image segmentation has become a standard methodology for medical image segmentation.  \cite{Ronneberger:2015gk}. It was further modified by the addition of residual layers in \cite{zhang2018road}.
For experiments presented here, we use the architecture shown in figure \ref{fig:model}.  Implementation  used the Keras \cite{chollet2015keras} interface to  the Tensorflow \cite{abadi2016tensorflow} library. 

Our network was designed to operate on 128x128 pixel patches of imaging data.  In our experiments we generated image patches from axial slices extracted from 3D data  using 64 px strides for spinal canals. For disks we used sagittal slices and perform the same patch extraction. Input patches are collected from pre-processed input scan(s) and output patches are collected from corresponding manual segmentation(s).  Before training the model, patches extracted from images were augmented by transforming each patch (and the corresponding segmentation) by a randomly picked combination of a translation, rotation and scaling. Specifically for each patch, the augmentation algorithm randomly picked an angle between $+/- 20^o$, a scaling factor between $[0.8,1.2]$ and x-translation and y-translation limited by $+/-50\;$ px.  For training models used in the Eigenrank selection process (see section F) each patch is augmented 20 times, since these models are based on small data subsets. \cite{gaonkar2018extreme}. For training models which are used to validate the Eigenrank selection procedure (see Results), each patch is augmented twice.

\subsection{Terminology}

 \begin{itemize}
 
\item We denote the training set as $\mathcal{T}$, the set of pairs $\lbrace (I_1, S_1) , (I_2, S_2), \cdots (I_j,S_j) \cdots (I_N,S_N)\rbrace$ with  $I_j$ representing $j^{th}$ patient scan and $S_j$ representing the corresponding segmentation image.

\item For any subset  $\mathcal{S} \subset \mathcal{T}$ we define $\mathcal{D}_{\mathcal{S}}$ as the deep learning model trained using a chosen deep learning segmentation algorithm denoted by $\mathcal{D}$

\item Further we denote by $S_j^{\mathcal{D}_{\mathcal{S}}}$ the segmentation image obtained by applying the model $\mathcal{D}_{\mathcal{S}}$ to the image $I_j \in \mathcal{T}$

\item The Dice coefficient of overlap is denoted by operator $\Delta(. ,.)$ So, the Dice coefficient comparing $S_j^{\mathcal{D}_{\mathcal{S}}}$ annd $S_j$ would be $\Delta(S_j^{\mathcal{D}_{\mathcal{S}}},S_j)$

\item Further we remind the reader of the set difference notation. Given sets $\mathcal{A}$ and $\mathcal{B}$ , the set $\mathcal{A}\setminus\mathcal{B}$ contains all elements of $\mathcal{A}$ that are not in $\mathcal{B}$

\end{itemize}

\subsection{The Eigenrank Algorithm}

The Eigenrank algorithm has an initialization phase and an iterative phase. The initialization step of Eigenrank closely follows the query-by-committee (QBC) \cite{seung1992query} paradigm although with the  modification that Dice coefficients used to affirm model agreement are real numbers rather than binary labels.  In the initialization phase the algorithm randomly selects two subsets of size $k$ from $\mathcal{T}$, trains deep learning segmentation models on these. Then it compares segmentations generated using one model to the other on the remanent of the training images using the Dice score. Note that this compares segmentations generated by one model to another and does not need ground truth. Images corresponding the lowest `$k$' Dice coefficients, are used to 'select' the next subset to train on.  The second step of Eigenrank is the `iterative' step, in which, we have to compare segmentation results from more than two deep learning models. This presents a unique problem which we solve by generating a Dice matrix.  At the $t^{th}$ iteration, $t$ models are available, each trained on a distinct $k-$ subset of $\mathcal{T}$.  We use these models to construct a $t \times t$ matrix, whose elements are Dice scores comparing segmentations derived from each model with every other model.  The  principal eigenvalue of this matrix  serves as a measure of disagreement among these $t$ models.  This principal eigenvalue is representative of the Von Neumann entropy of the  Dice matrix, a connection further elucidated in the Discussion section. Selecting  images corresponding to the minimum $k$ principal eigenvalues of the  Dice matrix takes us to the  ${t+1}^{th}$ iteration. We formally present the algorithm next.

\noindent\rule{9cm}{0.4pt} \\
\textbf{Initialization}\\
\noindent\rule{9cm}{0.4pt} 
\begin{itemize}[\relax]

\item From $\mathcal{T}$ randomly select subsets $\mathcal{S}_1,\mathcal{S}_2 \subset \mathcal{T}$ 

\begin{itemize}
\item[] we require   $k=|\mathcal{S}_1|=|\mathcal{S}_2|<<|\mathcal{T}|$
\item[] and define   $ \mathcal{S} \doteq \mathcal{S}_1 \cup \mathcal{S}_2$
\end{itemize}

\item Train U-Nets $\mathcal{D}_{\mathcal{S}_1}$ and $\mathcal{D}_{\mathcal{S}_2}$ and define $\mathcal{L}=\{\}$

\item For all $ I_j \in \mathcal{T}\setminus\mathcal{S}$,
\begin{itemize} 
\item Compute segmentations $S_j^{\mathcal{D}_{\mathcal{S}_1}}$ and $S_j^{\mathcal{D}_{\mathcal{S}_2}}$ \item  Compute Dice score $\Delta_j^{\mathcal{S}_1,\mathcal{S}_2} \doteq \Delta(S_j^{\mathcal{D}_{\mathcal{S}_1}},S_j^{\mathcal{D}_{\mathcal{S}_2}})$
\item $\mathcal{L}=\mathcal{L}\cup\{\Delta_j^{\mathcal{S}_1,\mathcal{S}_2}\}$
\end{itemize}
\item Use images corresponding to the $k-$smallest values in $\mathcal{L}$ to construct $\mathcal{S}_{3}$
\item Set $\mathcal{S}=\mathcal{S}\cup\mathcal{S}_3$
\end{itemize}
\noindent\rule{9cm}{0.4pt} \\
\textbf{Iterations}\\
\noindent\rule{9cm}{0.4pt} 
\begin{itemize}
\item For $t$ in $\{3, \cdots, T\}$
\begin{itemize}
\item[o] Train model $\mathcal{D}_{\mathcal{S}_t}$ and set $\mathcal{L}=\{\}$
\item[o] Using $\{\mathcal{D}_{\mathcal{S}_1}, \cdots \mathcal{D}_{\mathcal{S}_t}\}$ compute:
\begin{itemize}
\item[-] For  all$ \;\;I_j \in \mathcal{T}, I_j \not\in \mathcal{S}$
\begin{itemize}
\item []Compute $\; D_j^{pq}\doteq\Delta_j^{\mathcal{S}_p,\mathcal{S}_q} = \Delta(S_j^{\mathcal{D}_{\mathcal{S}_p}}$ , $S_j^{\mathcal{D}_{\mathcal{S}_q}})$ 
 
 \item[] $\forall p,q \in \{ 1, \cdots, t\} \;\; $
\item[] Define: $\;\mathbf{D}_j\doteq[D_j^{pq}] \in  \mathbb{R}^{t \times t}$
\item[]Compute $\;\lambda_j^{max}=max(eig[\mathbf{D}_j])$
\item[] $\mathcal{L}=\mathcal{L}\cup\{\lambda_j^{max}\}$
\end{itemize}
\end{itemize}
\item[o] Use images corresponding to the $k-$smallest values in $\mathcal{L}$ to construct $\mathcal{S}_{t+1}$
\item[o] Set $\mathcal{S}=\mathcal{S}\cup\mathcal{S}_{t+1}$\\
\end{itemize}
\end{itemize}
\noindent\rule{9cm}{0.4pt} \\
\textbf{Output}\\
\noindent\rule{9cm}{0.4pt} \\
Output  selected data subset $\mathcal{S}$ \\
\noindent\rule{9cm}{0.4pt} 


%
\subsection{Eigenrank for Failure Prediction}
The intuition that drives  Eigenrank  also provides a framework for failure prediction. If multiple models lack strong agreement over segmenting a particular scan, such a case is best referred to a human expert.  This process closely mimics  what human trainees do. The theory behind Eigenrank is based on a  framework that can quantify how much a group of image segmentation models disagree on a particular scan. Thus, it can be used to select scans which are likely to challenge deep learning based anatomy segmentation, and refer such cases to human experts.

\section{Results}
In this section we show both qualitative as well as quantitative results outlining how Eigenrank selects data subsets/predicts failure. We present empirical evidence about the effectiveness of Eigenrank. by comparing it with random selection.  For experiments shown here, we run Eigenrank with $k=3$ to `pick' data subsets for training two different models. The first model aims to segment spinal canals on axial MR images. The second aims to segment intervertebral disks on sagittal MR images.

\subsection{Eigenrank for spinal canal segmentation}
\subsubsection{Eigenrank for data subset selection}
 The first set of experiments used  100 axial T2-MRI scans of the lumbar spine, on which manual segmentations  of the spinal canal are available.  The scans used were randomly selected from a clinical imaging database. They contained artifacts due to variation in acquisition, pathology as well as metallic implants and surgical hardware often used in treating spine related conditions.  We expect a robust  segmentation algorithm to achieve accurate segmentation despite the presence of these artifacts.  Thus, a robust algorithm will have both high average dice score and a lower standard deviation in Dice scores. The more the robustness, the better the applicability to a clinical scenario.  We ran Eigenrank for 7 iterations thus choosing 21 images. We also selected 21 random images from the data in groups of 3.  
 
\begin{figure}
  \centering
  \includegraphics[width=0.95\linewidth]{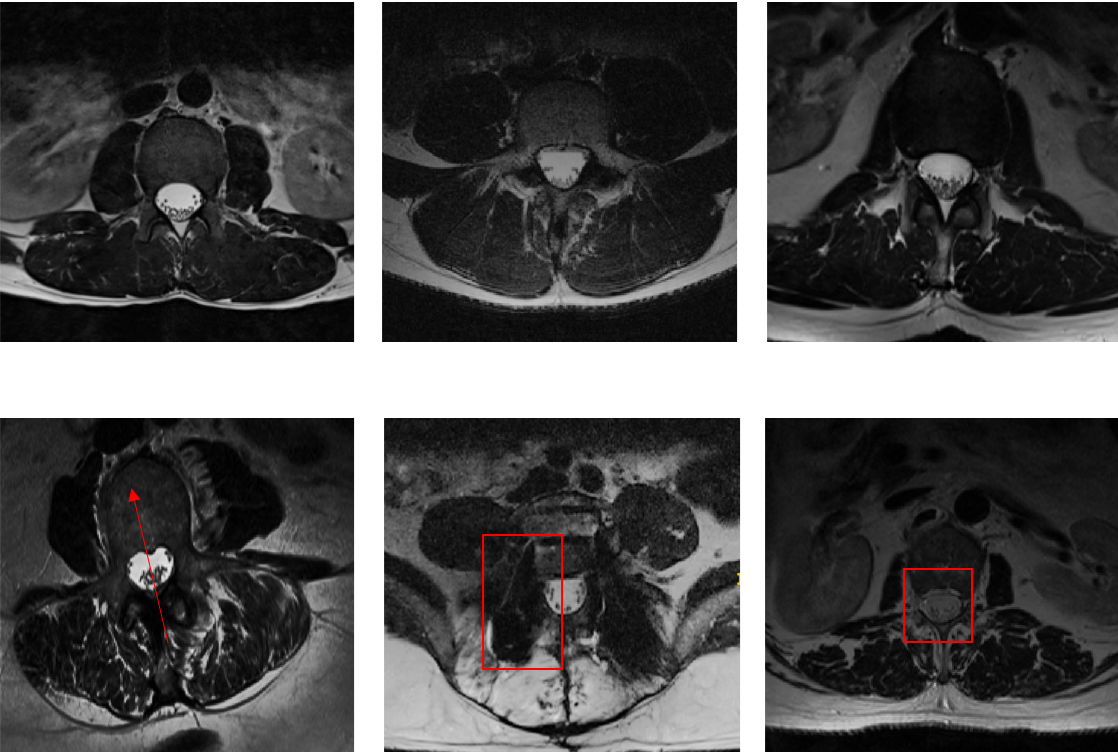}
  \caption{\textbf{Top Row} Randomly selected subjects \textbf{Bottom Row} Subjects selected by the first iteration of the proposed algorithm}
  \label{fig:selection}
\end{figure}

Three, cases selected via random sampling are presented alongside for comparison. It can be seen from the figure \ref{fig:selection} that Eigenrank selects cases which are much more complex as compared to random sampling.  One of these cases has abnormally scoliotic pathology, the second has screws and the third has a relatively lower contrast, perhaps due to an acquisition issue. By contrast the variation in both intensity, pathology and instrumentation within cases picked randomly is, distinctively lower. 

\begin{table}[]
\caption{Comparing models trained on data selected using Eigenrank at various iterations and Random selection on a fixed left-out validation set which excludes cases selected by Eigenrank while validating random models. Note that Eigenrank produces models which consistently have lower standard deviation - indicating higher robustness}
\label{tab:my-table}
\begin{center}
\begin{tabular}{|l|l|l|}
\hline
 & \begin{tabular}[c]{@{}l@{}}Eigenrank \end{tabular} & Random \\ \hline
 
\multicolumn{3}{|l|}{\textbf{Iteration} $t= 7$} \\ \hline
Mean Dice & 0.88 & 0.87 \\ \hline
Std. Dev Dice & \textbf{0.035} & 0.041 \\ \hline

\multicolumn{3}{|l|}{\textbf{Iteration} $t=6$} \\ \hline
Mean Dice & 0.87 & 0.87 \\ \hline
Std. Dev Dice & \textbf{0.038} & 0.041 \\ \hline

\multicolumn{3}{|l|}{\textbf{Iteration} $t=5$} \\ \hline
Mean Dice & 0.87 & 0.86 \\ \hline
Std. Dev Dice & \textbf{0.036} & 0.048 \\ \hline

\multicolumn{3}{|l|}{\textbf{Iteration} $t=4$} \\ \hline
Mean Dice & 0.85 & 0.85 \\ \hline
Std. Dev. Dice & \textbf{0.041} & 0.052 \\ \hline

\multicolumn{3}{|l|}{\textbf{Iteration} $t=3$} \\ \hline
Mean Dice & 0.85 & 0.85 \\ \hline
Std. Dev Dice & \textbf{0.042} & 0.052 \\ \hline
\end{tabular}
\end{center}
\end{table}

In the table \ref{tab:my-table}, we compare the mean and the standard deviation of the Dice score attained by U-Net models trained using Eigenrank to those trained by purely random selection for 7 iterations ($t=3$ to $t=7$).
After each iteration the data subset selected grows by 3. The entire data is used to train a residual U-Net. Patches extracted from the data are augmented only twice for these models. It can be seen from the table that Eigenrank selects data subsets which lead to  deep learning models for spinal canal image segmentation which present a lower standard deviation on the 60 validation cases, while maintaining high accuracies.  
\textbf{The lower standard deviations show that models trained on data selected using Eigenrank are more robust than those selected using random selection, and slightly more accurate as well.} This is very important when one considers clinical deployment of deep learning models, where robustness is often as important as  accuracy. We also note that, the numbers presented in table \ref{tab:my-table}   are generated using a set of data that do not include either cases selected by Eigenrank or random cases. Since, Eigenrank weeds out all the complex cases in the data, the random model is effectively validated against clean data, making the difference between Eigenrank and Random selection less extreme.

In real clinical settings if Eigenrank is not available to extract challenging cases and ensure that human physicians segment them, we would have a scenario where a model trained on randomly selected data is used on the remaining available data for diagnostic or biomarker extraction purposes. To mimic this case it would be more appropriate to compute Dice scores on data left-out by Eigenrank as well as random selection after each iteration, rather than a common dataset left unselected by both methods. This comparison is presented in table \ref{tab:dssc}.

\begin{table}[]
\caption{Comparing models trained on data selected using Eigenrank  at various iterations and Random selection on a validation sets which only exclude images selected by the specific algorithm at a specific iteration. So for instance at $t=4$ with Eigenrank, the training dataset contains 12 cases and validation contains 88. Again, Eigenrank produces models which consistently have lower standard deviation - indicating higher robustness}
\label{tab:dssc}
\begin{center}
\begin{tabular}{|l|l|l|}
\hline
 & \begin{tabular}[c]{@{}l@{}}Eigenrank \end{tabular} & Random \\ \hline
 
\multicolumn{3}{|l|}{\textbf{Iteration} $t= 7$} \\ \hline
Mean Dice & 0.88 & 0.84 \\ \hline
Std. Dev Dice & \textbf{0.036} & 0.11 \\ \hline

\multicolumn{3}{|l|}{\textbf{Iteration} $t=6$} \\ \hline
Mean Dice & 0.87 & 0.84 \\ \hline
Std. Dev Dice & \textbf{0.054} & 0.123 \\ \hline

\multicolumn{3}{|l|}{\textbf{Iteration} $t=5$} \\ \hline
Mean Dice & 0.87 & 0.83 \\ \hline
Std. Dev Dice & \textbf{0.055} & 0.123\\ \hline

\multicolumn{3}{|l|}{\textbf{Iteration} $t=4$} \\ \hline
Mean Dice & 0.84 & 0.82 \\ \hline
Std. Dev. Dice & \textbf{0.067} & 0.126 \\ \hline

\multicolumn{3}{|l|}{\textbf{Iteration} $t=3$} \\ \hline
Mean Dice & 0.84 & 0.82 \\ \hline
Std. Dev Dice & \textbf{0.074} & 0.126 \\ \hline
\end{tabular}
\end{center}
\end{table}

\subsubsection{Failure analysis using Eigenrank}
Could Eigenrank serve as a method for predicting 'failure' of a deep learning models. In this section we present an experiment which supports this case. We train a deep learning model $D_m$ on a random subset of $15$ scans. This leaves $85$ scans for validation. We eliminate scans from these 85 in batches of $k=3$ using Eigenrank. So after the first iteration, $82$ validation cases remain and $3$ cases are eliminated.  We use $D_m$ to segment the validation set and compare the resulting segmentation to  compute the average Dice score on the $82$ cases as well as the $3$ remaining scans .  As Eigenrank eliminates the complex cases, the average Dice score on the remaining cases increases and the standard deviation of the Dice scores decreases. This indicates that Eigenrank can preemptively detect cases which the deep residual U-Net model will find challenging.

\begin{table}[]
\caption{Using Eigenrank purely as a failure analysis method for \textbf{spinal canal segmentation}. A model is created  by training on a  \textbf{left-out} set of 15 cases. Out of the remaining 85 validation cases, Eigenrank was used to iteratively remove 'difficult' cases. After each iteration, we compute the mean Dice score of the cases eliminated and of the remaining validation cases}
\label{tab:my-table}
\begin{center}
\begin{tabular}{|l|l|l|l|l|}
\hline
\textbf{Iteration} & \multicolumn{2}{l|}{\textbf{\begin{tabular}[c]{@{}l@{}}For cases\\ eliminated by\\ Eigenrank\end{tabular}}} & \multicolumn{2}{l|}{\textbf{\begin{tabular}[c]{@{}l@{}}For  remaining\\ cases\end{tabular}}} \\ \hline
 & \begin{tabular}[c]{@{}l@{}}Mean\\ Dice\end{tabular} & \begin{tabular}[c]{@{}l@{}}Stdev\\ Dice\end{tabular} & \begin{tabular}[c]{@{}l@{}}Mean\\ Dice\end{tabular} & \begin{tabular}[c]{@{}l@{}}Stdev\\ Dice\end{tabular} \\ \hline
1 & 0.611 & 0.317 & 0.835 & 0.076 \\ \hline
2 & 0.650 & 0.265 & 0.839 & 0.075 \\ \hline
3 & 0.710 & 0.253 & 0.840 & 0.075 \\ \hline
4 & 0.742 & 0.238 & 0.840 & 0.076 \\ \hline
5 & 0.750 & 0.222 & 0.842 & 0.074 \\ \hline
6 & 0.769 & 0.212 & 0.840 & 0.074 \\ \hline
7 & 0.749 & 0.208 & 0.851 & 0.052 \\ \hline
\end{tabular}
\end{center}
\end{table}

\subsection{Eigenrank for intervertebral disk segmentation}
\begin{figure}
  \centering
  \includegraphics[width=0.95\linewidth]{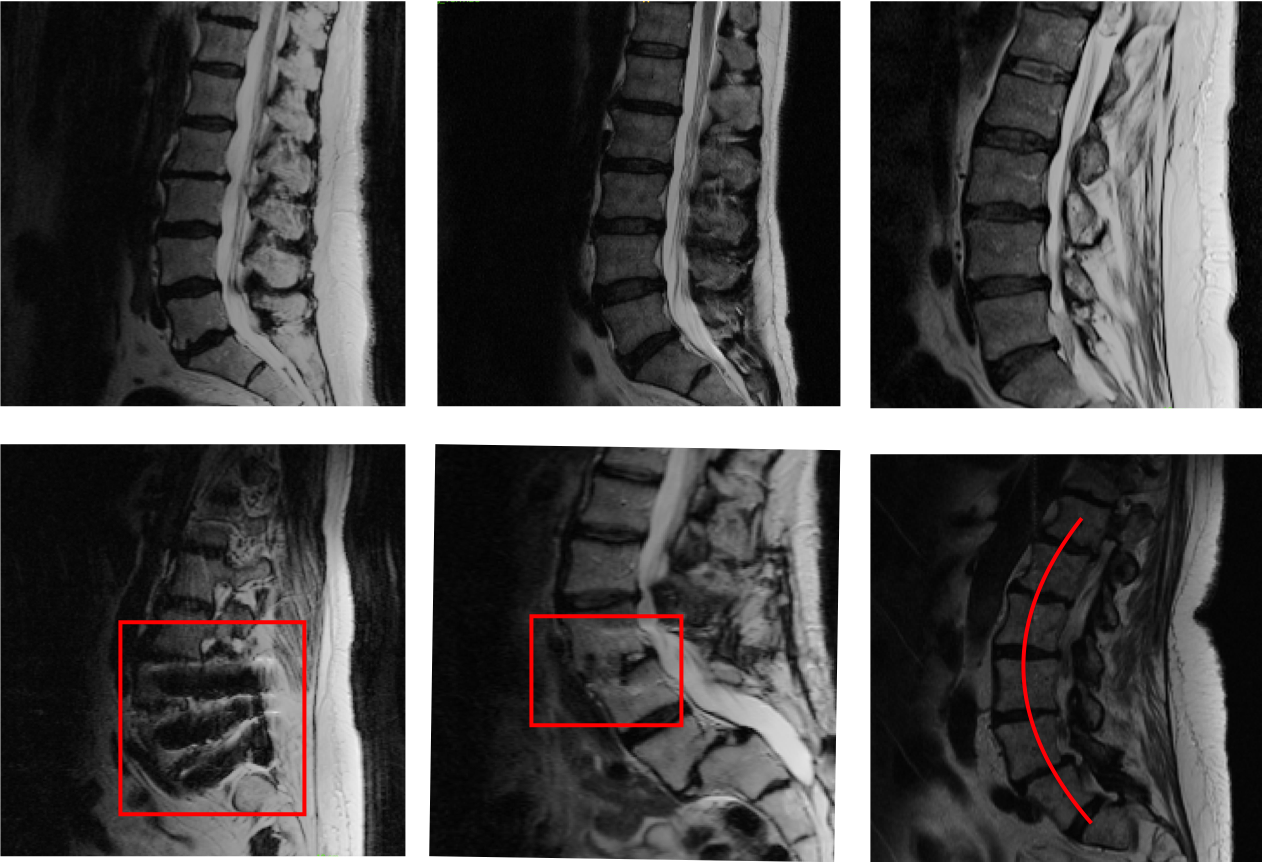}
  \caption{\textbf{Top Row} Randomly selected subjects \textbf{Bottom Row} Subjects selected by the first iteration of the Eigenrank}
  \label{eigen:disk}
\end{figure}

To further convince the reader of the value of Eigenrank, we ran the aforementioned experiments in a separate dataset where the objective was automatically segmenting disks on  lumbar spine MR images. The full dataset contained 103 MR scans. Inter-vertebral disks were segmented manually using techniques similar to spinal canals. We ran Eigenrank using same network and parameters as the previous experiment. A sample of images selected by Eigenrank are shown in figure \ref{eigen:disk}. Standard deviations and means of  Dice scores associated with data subset selection on left-out data are shown in table \ref{dss:disk}. Dice scores on validation data derived by eliminating `cases chosen' by each specific algorithm iteratively are shown in table \ref{tab:dssd}. Failure prediction results are shown in \ref{fp:disk}. 

\begin{table}[]
\caption{Comparing models trained on data selected using Eigenrank at various iterations and Random selection. Note that Eigenrank produces models which consistently have lower standard deviation - indicating higher robustness}
\label{dss:disk}
\begin{center}
\begin{tabular}{|l|l|l|}
\hline
 & \begin{tabular}[c]{@{}l@{}}Eigenrank \end{tabular} & Random \\ \hline
 
\multicolumn{3}{|l|}{\textbf{Iteration} $t= 7$} \\ \hline
Mean Dice & 0.86 & 0.86 \\ \hline
Std. Dev Dice & \textbf{0.032} & 0.036 \\ \hline

\multicolumn{3}{|l|}{\textbf{Iteration} $t=6$} \\ \hline
Mean Dice & 0.85 & 0.85 \\ \hline
Std. Dev Dice & \textbf{0.032} & 0.038 \\ \hline

\multicolumn{3}{|l|}{\textbf{Iteration} $t=5$} \\ \hline
Mean Dice & 0.85 & 0.84 \\ \hline
Std. Dev Dice & \textbf{0.035} & 0.039 \\ \hline

\multicolumn{3}{|l|}{\textbf{Iteration} $t=4$} \\ \hline
Mean Dice & 0.84 & 0.84 \\ \hline
Std. Dev. Dice & \textbf{0.035} & 0.039 \\ \hline

\multicolumn{3}{|l|}{\textbf{Iteration} $t=3$} \\ \hline
Mean Dice & 0.81 & 0.81 \\ \hline
Std. Dev Dice & \textbf{0.037} & 0.039 \\ \hline
\end{tabular}
\end{center}
\end{table}

\begin{table}[]
\caption{Comparing models trained on data selected using Eigenrank at various iterations and Random selection on a validation set which excludes data selected by a particular algorithm. Again, Eigenrank produces models which consistently have lower standard deviation - indicating higher robustness}
\label{tab:dssd}
\begin{center}
\begin{tabular}{|l|l|l|}
\hline
 & \begin{tabular}[c]{@{}l@{}}Eigenrank \end{tabular} & Random \\ \hline
 
\multicolumn{3}{|l|}{\textbf{Iteration} $t= 7$} \\ \hline
Mean Dice & 0.86 & 0.84 \\ \hline
Std. Dev Dice & \textbf{0.032} & 0.065 \\ \hline

\multicolumn{3}{|l|}{\textbf{Iteration} $t=6$} \\ \hline
Mean Dice & 0.85 & 0.84 \\ \hline
Std. Dev Dice & \textbf{0.032} & 0.066 \\ \hline

\multicolumn{3}{|l|}{\textbf{Iteration} $t=5$} \\ \hline
Mean Dice & 0.85 & 0.83 \\ \hline
Std. Dev Dice & \textbf{0.037} & 0.067 \\ \hline

\multicolumn{3}{|l|}{\textbf{Iteration} $t=4$} \\ \hline
Mean Dice & 0.84 & 0.83 \\ \hline
Std. Dev. Dice & \textbf{0.036} & 0.069 \\ \hline

\multicolumn{3}{|l|}{\textbf{Iteration} $t=3$} \\ \hline
Mean Dice & 0.83 & 0.82 \\ \hline
Std. Dev Dice & \textbf{0.037} & 0.068 \\ \hline
\end{tabular}
\end{center}
\end{table}

\begin{table}[]
\caption{ Using Eigenrank as a failure analysis method for \textbf{disk segmentation on saggital MRI}. A model to segment disks in saggital MRI is created  by training on a  \textbf{left-out} set of 15 cases. Out of the remaining 87 validation cases, Eigenrank was used to iteratively remove `difficult' cases. After each iteration, we compute the mean Dice score of the cases eliminated and of the remaining validation cases}
\label{fp:disk}
\begin{center}
\begin{tabular}{|l|l|l|l|l|}
\hline
\textbf{Iteration} & \multicolumn{2}{l|}{\textbf{\begin{tabular}[c]{@{}l@{}}For cases\\ eliminated by\\ Eigenrank\end{tabular}}} & \multicolumn{2}{l|}{\textbf{\begin{tabular}[c]{@{}l@{}}For  remaining\\ cases\end{tabular}}} \\ \hline
 & \begin{tabular}[c]{@{}l@{}}Mean\\ Dice\end{tabular} & \begin{tabular}[c]{@{}l@{}}Stdev\\ Dice\end{tabular} & \begin{tabular}[c]{@{}l@{}}Mean\\ Dice\end{tabular} & \begin{tabular}[c]{@{}l@{}}Stdev\\ Dice\end{tabular} \\ \hline
1 & 0.681 & 0.154 & 0.835 & 0.040 \\ \hline
2 & 0.739 & 0.152 & 0.839 & 0.040 \\ \hline
3 & 0.773 & 0.145 & 0.840 & 0.040 \\ \hline
4 & 0.789 & 0.135 & 0.840 & 0.040 \\ \hline
5 & 0.804 & 0.132 & 0.842 & 0.040 \\ \hline
6 & 0.805 & 0.125 & 0.840 & 0.038 \\ \hline
7 & 0.809 & 0.119 & 0.851 & 0.037 \\ \hline
\end{tabular}
\end{center}
\end{table}

\section{Discussion}
We have presented our algorithm from a utilitarian point of view. In this section we first present  intuitions which drove the design of Eigenrank.  Then, we discuss alternative metrics which could be used in Eigenrank, in place of the  eigenvalue measure proposed. We also discuss in detail why we consider Eigenrank a better alternative to traditional data subset selection in medical image segmentation. We also highlight how Eigenrank is related to QBC and note some of the mathematical problems which emerge from our experiments.

\subsection{Von Neumann Information}
The Von Neumann entropy can be defined for the symmetric positive definite matrix  can be defined to be the sum of the Shannon entropy of it's Eigenvalues. The matrix $\mathbf{D}_j$ can be proved to be positive semi-definite \cite{nader2019positive}. The proof follows from the fact that we can express:
\begin{equation}
 D_j^{pq}= \frac{2 \mathbf{s}_p . \mathbf{s}_q}{|\mathbf{s}_p|+|\mathbf{s}_q|}
\end{equation}
where:
\begin{equation}
\mathbf{s}_p,\mathbf{s}_q \in \lbrace 0,1 \rbrace^{|I_j|}
\end{equation}
are the vectorized representations of images $S_j^{\mathcal{D}_{\mathcal{S}_p}}$ , $S_j^{\mathcal{D}_{\mathcal{S}_q}}$
Thus, the Dice matrix itself can be thought of as a Hadamard product of an inner product matrix and a Cauchy matrix. That is:
\begin{equation}
\mathbf{D}_j = 2 \mathbf{K}_j \circ \mathbf{C}_j
\end{equation}
where we define:
\begin{equation}
{K}^{pq}_j = \mathbf{s}_p . \mathbf{s}_q 
\end{equation}
and
\begin{equation}
{C}^{pq}_j=\frac{1}{|\mathbf{s}_p|+|\mathbf{s}_p|}
\end{equation}
Thus,  $\mathbf{K}_j$ is an inner product matrix - which are always positive semi-definite. The Cauchy matrix  is positive semi-definite \cite{bhatia2009positive} because it can be expressed as an inner product matrix in Hilbert space:
\begin{equation}
\frac{1}{|\mathbf{s}_p|+|\mathbf{s}_q|}=\int_0^\infty e^{-{(|\mathbf{s}_p|+|\mathbf{s}_q|)t}}dt
\end{equation}
and
\begin{equation}
\int_0^\infty e^{-{(|\mathbf{s}_p|+|\mathbf{s}_q|)t}}dt=\int_0^\infty e^{-{|\mathbf{s}_p|t}}.e^{-{|\mathbf{s}_q|t}}dt
\end{equation}

Given that both $\mathbf{K}_j$ and $\mathbf{C}_j$ is positive semi-definite the Schur product theorem then ensures that $\mathbf{D}_j$ is positive semi-definite as well. The proof presented here is based on previous work by Nader \cite{nader2019positive}  and Bhatia \cite{bhatia2009positive} which the reader should refer to for more details.
 In Eigenrank we expect each of the $t$ models to generate unique segmentations and expect $\mathbf{D}_j$ to be positive definite rather than positive semi-definite and we define the associated Von Neumann entropy as:
\begin{equation}
H_j=-\sum_{r=1 \cdots t  } \lambda_r log(\lambda_r) 
\end{equation}
where $\lbrace \lambda_1 > \lambda_2 >\cdots > \lambda_t \rbrace$ are the ordered eigenvalues of $\mathbf{D}_j$. 


In the experiments described here it generally turns out that,  $\lambda_1 log(\lambda_1)$ dominates $H_j$ because:
\begin{equation}
\label{eq:expect}
\lambda_1>> \lambda_2,\;\lambda_3,\;\lambda_4, \cdots \;\lambda_t
\end{equation}
Thus, we intuit that Eigenrank effectively looks for cases with the highest Von Neumann entropy. To understand why we can generally expect (\ref{eq:expect}) to be true, consider 
the two extremes of $\mathbf{D}_j=\mathbf{I}$ with $\mathbf{I} \in \mathbb{R}^{t \times t}$ and $\mathbf{D}_j=\mathbf{J}$ with $\mathbf{J} \in \mathbb{R}^{t \times t}$. Both of these cases never occur inn practice but correspond to specific fictional scenarios. When $\mathbf{D}_j =\mathbf{I}$,  each model agrees with itself but disagrees completely with every other model. When $\mathbf{D}_j =\mathbf{J}$ all models fully agree with each other. In the first case all eigenvalues of $\mathbf{D}_j$ are unity.  In the second case, we can analytically work out $\lambda_1=t$ and $\lambda_2,\cdots \lambda_t =0$ and (\ref{eq:expect}) will be true and the maximum eigenvalue dominates the Von Neumann information.  In most cases of practical interest, we would expect the various models involved to \textit{mostly but not completely} agree and the maximum eigenvalue remains an effective metric. 
The analysis is presented to give the reader a perspective into why the proposed metric for ordering scans works by linking it to a well established information theoretic concept. However, we believe that Eigenrank is a more general framework and could potentially be utilized with other metrics  as well. We discuss these alternate possibilities next. 
\subsection{Alternate Metrics}
First, Eigenrank is based on generalizing the concept of `disagreement' between more than two image segmentation models on a global image wide rather than a local pixel/patch level. While the Dice coefficient defines disagreement in a relatively straightforward manenr, when comparing two segmentations, it is unclear how to compare more than two segmentations and segmentation models amongst themselves. Eigenrank relies on a specific generalization of the Dice coefficient to more than two segmentations. However, the Dice coefficient is not the only available metric for comparing two segmentations. The Jaccard index, the Hausdorff distance and the average surface/contour distance are all valid and accepted metrics for this purpose. These metrics could potentially be utilized in alternative versions of Eigenrank-like algorithms. While a thorough investigation of each metric is out of scope in the present manuscript, we nevertheless record each metric and discuss it with respect to what has already been proposed. The first of these metrics is the Jaccard index which could be used directly in the proposed framework. It is interesting to note that for both the Jaccard index and the Dice score can be 
 generalized directly to compare three or more segmentations per case. However, such generalizations yield metrics which become zero if even one model in the ensemble $\{\mathcal{D}_{\mathcal{S}_1}, \cdots \mathcal{D}_{\mathcal{S}_t}\}$ fails completely or generates a segmentation which has no overlap with any of the other segmentations.  Hausdorff distance could be used but it lacks the nice connection with Von Neumann entropy described in the last section. One would have to understand how pairwise Hausdorff distances can be compiled into a matrix, and what measure derived from such a matrix could be used for data selection. The same would be true for average surface distances. \textit{The full Von Neumann information could potentially provide a nicer metric than maximum eigenvalue to use within Eigenrank, but we verified that the ultimate results of selection and failure prediction remain identical.}

\subsection{Relationship to Query-by-Committee}
 The query-by-committee (QBC) framework of active learning, first presented by Seung \cite{seung1992query} motivated Eigenrank. QBC, operates on a framework similar to Eigenrank, where multiple models are trained on current labels, and  new candidates for training are picked based on where  "the committee" disagrees the most. QBC was first proposed from an information theoretic perspective , and further developed in it \cite{freund1997selective}. Later, other authors extended QBC with kernels \cite{gilad2006query} and studied its theoretical properties \cite{buus2003sensitive}.The  premise of QBC based DSS is that  a data instance which two machine learning models label differently, is more informative for the training subset. In the standard classification, setting where labels are either binary or discrete, this premise is straightforward to apply. However, in deep learning based medical image segmentation, applying QBC directly presents several challenges unless one is applying it at a pixel level where standard deviation of segmentation intensities provides a simple metric to quantify disagreement. Applying QBC at the scan level requires that comparisons of outputs of multiple models are made at the global segmented image level. Eigenrank presents one paradigm in which this may be done.

\subsection{Comparison to traditional data subset selection}
Traditional data subset selection methods (as well as active learning) algorithms have most often been designed for either classification or regression problems. Their applicability in medical image segmentation is thus fairly limited. Further most traditional data subset selection techniques tend to operate independently of the algorithm. For instance, facility location based submodular dataset selection, would select the same subset whether, we were segmenting a spinal canal or  spinal vertebrae or some other anatomical structure. Eigenrank, on the other hand, has the potential adapt  selection and  selection strategy to the specific anatomical substructure of interest. This is true of active learning in general. Yet, the majority of literature on active learning for medical image segmentation focuses on identifying variance between models  at a local pixel/voxel level rather than a global entire image level. It is unclear whether such disagreement at the pixel or patch level translates to overall disagreement at the scan level. Moreover, it is easy to imagine  scenarios where local disagreement  does not translate to global disagreement. For instance, the existence of an unusually bright pixel, may cause certain models to fail locally causing local variance, yet globally a single pixel being mis-segmented hardly matters. To the best of our knowledge Eigenrank based data subset selection is unique in quantifying and utilizing inter-model variance on an full image basis for data subset selection and active learninng. This global variance quantification using Von Neumann entropy places Eigenrank uniquely in the space of active learning methods used in medical image analysis.

\subsection{A note of model selection}
In this work we have used a specific instance of a residual U-Net model to both construct and validate our framework.Perhaps a completely different model,  patch generation and data augmentation scheme could  be used. As such, hyper-parameter optimization, learning rate optimization, batch normalization, architecture optimization and all the other techniques which can improve deep networks could be used to create better models.However, our aim in this work is not to focus on model optimization, but rather to highlight the effectiveness of Eigenrank for data subset selection and failure prediction, rather than delving into theory behind deep learning. Hence, we used a relatively straightforward architecture with fixed hyperparameters, patch generation and augmentation schema.

\subsection{Mathematical aspects}
It is useful to understand the positive semi-definiteness of $\mathbf{D}_j\succeq 0$ from a geometric standpoint. Specifically we explore the implications of this for comparing three segmentations to each other.

\begin{figure}
  \centering
  \includegraphics[width=0.95\linewidth]{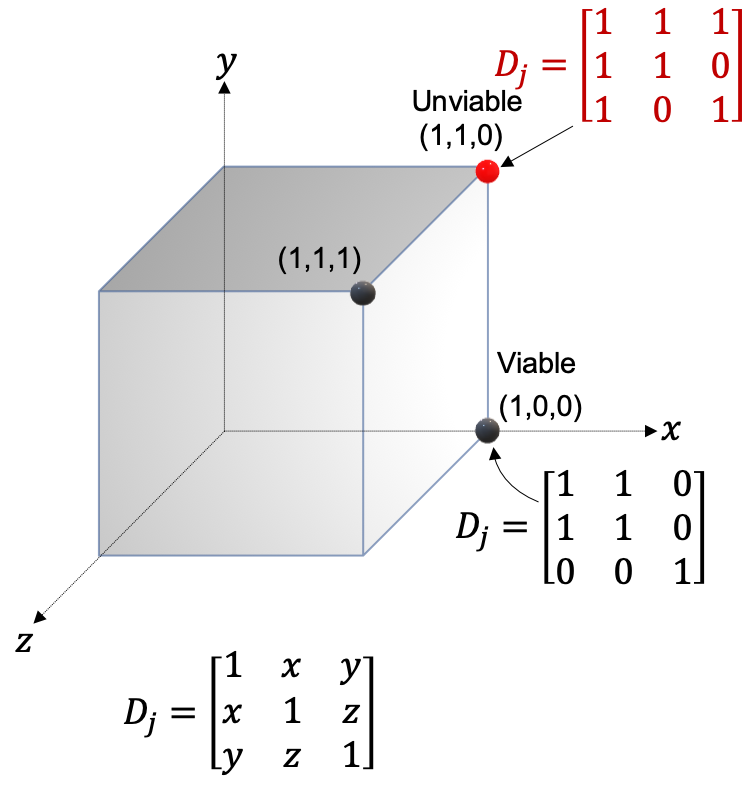}
  \caption{Viability and unviability of $\mathbf{D}_j$ in comparing three models to each other.}
  \label{c1}
\end{figure}

In the case presented by figure \ref{c1} it is possible to visualize why $\mathbf{D}_j$ this might be true for the relatively simple case of three models shown in figure \ref{c1}. In the case of three models, if two models agree with the third one, they cannot disagree among themselves. This unviable situation would lead to a non positive definite matrix 

\begin{center}
$\mathbf{D}_j = 
\left[ {
\begin{matrix} 
1 & 1 & 1 \\ 
1 & 1 & 0 \\ 
1 & 0 & 1  
\end{matrix}
 } \right]
$
\end{center}
with eigenvalues $[2.42,1,-0.42]$. Thus, the intersection of the cube and the cone of positive semi-definite  $\mathbf{D}_j$ forms a region of space where feasible Dice matrices arise.  $\mathbf{D}_j\succeq 0$  also leads to an elegant relationship between  dice coefficients arising out of mutual comparisons of segmentations generated by a trio of deep learning models. If $D^{pq}_j$, $D^{qr}_j$ and $D^{rp}_j$ are dice scores comparing segmentations generated on image $I_j$ by a trio of models $\mathcal{D}_{\mathcal{S}_p},\mathcal{D}_{\mathcal{S}_q},\mathcal{D}_{\mathcal{S}_r}$. Then, Conjecture 1 implies:
\begin{equation}
\label{constraint}
[D^{pq}_j]^2 + [D^{qr}_j]^2 + [D^{rp}_j]^2-1 < 2 D^{pq}_j D^{qr}_j D^{rp}_j
\end{equation}
This follows from the fact that the Schur complement of positive semi-definite matrix is positive semi-definite under the appropriate conditions. This can be used as an efficient testing criterion for simulating viable Dice matrices. We use it to test the following conjecture:

\textit{Conjecture: As the number of models $t$ increases the Shannon information of the maximum eigenvalue $\mathbf{D}_j$  dominates the Shannon information of all other eigenvalues.}

\begin{figure}
  \centering
  \includegraphics[width=0.95\linewidth]{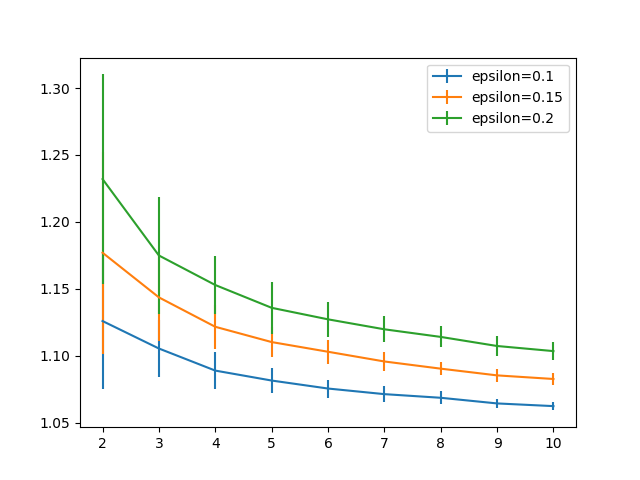}
  \caption{Why the largest eigenvalue of $\mathbf{D}_j$ suffices as a measure of disagreement for Eigenrank}
  \label{c2}
\end{figure}

If $\lambda_1 > \lambda_2 > \lambda_3 \cdots \lambda_t$ were sorted eigenvalues of $\mathbf{D}_j$ then this conjecture can be expressed as:
\begin{equation}
\lim_{t\to\infty}\frac{\lambda_1 log (\lambda_1)}{\sum_{r=1}^t \lambda_r log(\lambda_r)}=1
\end{equation}
 In Figure \ref{c2} we provide the results of simulations performed using randomly generated positive semi-definite matrices confirming to diagonal elements being equal to `1' and off-diagonal elements modeled as $1-\delta^{pq}$. Trios of $1-\delta^{pq}$ are constrained by (\ref{constraint}). $\delta^{pq}$ is randomly selected from the interval $[0,\epsilon]$ with $\epsilon$ set to various values. These simulations support the conjecture and this conjecture is the link connecting Eigenrank and information theory. Specifically it justifies the use of the maximum eigenvalue measure. Future work to ascertain the exact conditions under which it remains true, will be necessary to understand the limits of the proposed algorithm.

\section{Conclusion}
In conclusion, we have proposed a method for addressing both data subset selection and failure prediction, for deep learning based image segmentation.  We have also demonstrated the effectiveness of the proposed paradigm in two medical image analysis datasets. Our technique can help select subsets of images from large databases, in a manner such that accurate and more importantly `robust' deep neural networks can be trained for anatomical segmentation. It can also accurately identify challenging cases from a given dataset, where human attention is most likely needed. This gives deep learning based segmentation algorithms the ability to prioritize challenging cases within automated clinical image analysis workflows, thereby enabling better integration between human and machine in the future.

\section{Acknowledgment}
We thank the National Institutes of Health of the United States of America for support through the R21 grant mechanism. This work was funded by the grant R21EB026665.

\bibliographystyle{IEEETrans.bst}
\bibliography{IEEE.bib}

\end{document}